\documentclass[sigplan,nonacm]{acmart}
\acmSubmissionID{992}
\AtBeginDocument{%
  }

\setcopyright{acmlicensed}

\setcopyright{cc}
\setcctype[4.0]{by}

\makeatletter
\g@addto@macro\@copyrightpermission{%
  \par\vspace{.25\baselineskip}%
  \begingroup
    \setlength{\parskip}{0pt}\setlength{\parindent}{0pt}%
    \raggedright\footnotesize
    \textcopyright~2025 Copyright held by the owner/author(s). \\
    This is the author's version of the work. The definitive Version of Record was published in
    Proceedings of the Twenty-First European Conference on Computer Systems (EuroSys ’26),
    April 27--30, 2026, Edinburgh, Scotland UK, \href{https://doi.org/10.1145/3767295.3769374}{https://doi.org/10.1145/3767295.3769374}.%
    \par
  \endgroup
}
\makeatother

\usepackage{graphicx}
\usepackage{subcaption}
\usepackage{caption}
\usepackage{cleveref}
\usepackage{lipsum}
\usepackage{amsmath}
\usepackage{mathrsfs}
\usepackage{color,soul}
\usepackage{enumitem}
\usepackage{pifont}
\usepackage{ragged2e}
\usepackage{algorithm}
\usepackage{algpseudocode}
\usepackage{booktabs}
\usepackage{xcolor}
\usepackage{ulem}

\begin{document}

\newcommand{\redstrike}[1]{\textcolor{red}{\sout{#1}}}
\newcommand{\bluetext}[1]{\textcolor{blue}{#1}}
\newcommand{\orangetext}[1]{\textcolor{orange}{#1}}

\newcommand{\browntext}[1]{\textcolor{olive}{#1}}

\newcommand{\ms}[1]{\textcolor{magenta}{[Mohammad]: #1}}
\newcommand{\cl}[1]{\textcolor{orange}{[Changyuan]: #1}}

\title{Demystifying Serverless Costs on Public Platforms: Bridging Billing, Architecture, and OS Scheduling}



\author{Changyuan Lin}
\orcid{0000-0003-1618-9816}
\affiliation{
    \institution{University of British Columbia}
    \country{}}

\author{Yuanzhi Ma}
\orcid{0009-0000-3788-9956}
\authornote{Conducted the research while at The University of British Columbia.}
\affiliation{
    \institution{Johns Hopkins University}
    \country{}}

\author{Mohammad Shahrad}
\orcid{0000-0002-8214-9583}
\affiliation{
    \institution{University of British Columbia}
    \country{}}

\begin{abstract}
Public cloud serverless platforms have attracted a large user base due to their high scalability, plug-and-play deployment model, and pay-per-use billing. 
However, compared to virtual machines and container hosting services, modern serverless offerings typically impose higher per-unit time and resource charges.
Additionally, billing practices such as wall-clock time allocation-based billing, invocation fees, and usage rounding up can further increase costs. 

This work, for the first time, holistically demystifies these costs by conducting an in-depth, top-down characterization and analysis from user-facing billing models, through request serving architectures, and down to operating system scheduling on major public serverless platforms.
We quantify, for the first time, how current billing practices inflate billable resources up to $4.35 \times$ beyond actual consumption.
Also, our analysis reveals previously unreported cost drivers, such as operational patterns of serving architectures that create overheads, details of resource allocation during keep-alive periods, and OS scheduling granularity effects that directly impact both performance and billing.
By tracing the sources of costs from billing models down to OS scheduling, we uncover the rationale behind today's expensive serverless billing model and practices and provide insights for designing performant and cost-effective serverless systems.
\end{abstract}



\maketitle

\section{Introduction}
\label{sec:introduction}
Serverless computing has become one of the mainstream cloud computing paradigms, enabling developers to quickly deploy scalable and event-driven applications on the cloud without needing to manage the underlying infrastructure~\cite{jonas2019cloud, web:datadog23}.
Major cloud providers offer serverless computing solutions, such as AWS Lambda~\cite{web:lambda}, Google Cloud (GCP) Run functions~\cite{web:gcp}, Azure Functions~\cite{web:azure}, IBM Cloud Code Engine~\cite{web:ibm}, and Cloudflare Workers~\cite{web:cloudflare}.
Serverless computing stands out as the purest existing pay-per-use cloud model, offering automated scaling—from zero to thousands of instances in seconds—and fine-grained billing.
As a result, it is often advertised as cost-efficient~\cite{eismann2021state, web:lambda-pricing, jonas2019cloud, web:lambda-cost-efficient, web:serverless-cost-efficient}.

The widely acknowledged benefits of serverless architectures—such as high scalability, fine-grained pay-per-use billing, freedom from infrastructure management, and seamless integration with other cloud services—are not without associated costs~\cite{eivy2017wary, hellerstein2018serverless, wen2023rise}.
In terms of the per-unit resource price, serverless offerings are often priced higher than other cloud computing paradigms, such as virtual machines (VMs) and containers running on container hosting platforms.
We demonstrate this by comparing the price of AWS Lambda functions, AWS EC2 VMs, and AWS Fargate containers, all configured on identical ARM-based hardware in the \textit{us-east-2} region. 
We specifically chose ARM due to the diverse and performance-varying nature of AWS's x86 processors, which complicates fair comparisons across services.
An AWS Lambda function with 1\,vCPU, 1,769\,MB of memory, and 512\,MB of ephemeral storage costs $\$2.3034\times10^{-5}$ per second~\cite{web:lambda-pricing}, while a compute-optimized EC2 instance (\texttt{c6g.medium}) with 1\,vCPU, 2\,GB memory, and 1\,GB storage and an AWS Fargate container with the identical resource allocation as EC2 cost only $\$9.4753\times10^{-6}$ and $\$1.1003\times10^{-5}$ per second, which are 41.1\% and 47.8\% of the AWS Lambda price.
The cost of VMs can be further decreased by at least two times if using a burstable instance (e.g., AWS EC2 \texttt{t4g.small} flavor).
Also, this comparison does not include the invocation fee of AWS Lambda, which is $\$2\times10^{-7}$ for each request, whereas EC2 instances and Fargate containers do not charge request fees.
Additionally, our analysis of billing practices on major serverless platforms uncovers significant over-accounting (§\ref{sec:billing-practices}), showing that users can be charged for computing resources up to $4.35$ times greater than their actual usage.

These observations motivate a fundamental research question: \textbf{What makes serverless expensive?}
We argue that the root cause of the high unit prices and expensive billing practices in serverless lies in the architecture of modern serverless computing systems.
Resource consumption and overhead incurred by the underlying runtimes and control plane for request serving, such as sandbox provisioning, isolation, request dispatch, and keep-alive, translate into higher per-unit charges passed on to serverless users.
Additionally, some of our measurements of resource allocation patterns and performance behaviors on major serverless platforms point part of the execution costs and performance fluctuations to the underlying operating system (OS) scheduling mechanisms.

\setlength{\textfloatsep}{5pt}
\begin{table*}[h]
\resizebox{2\columnwidth}{!}{%
\begin{tabular}{|c|c|c|c|c|}
\hline
\textbf{Serverless Platform} &
  \textbf{Billable Time} &
  \textbf{Billable Resources$^*$} &
  \textbf{Billing Granularity/Cutoffs} &
  \textbf{Control Knobs and Steps} \\ \hline
AWS Lambda~\cite{web:lambda, web:aws-cpu-mem, web:lambda-pricing} &
  \begin{tabular}[c]{@{}c@{}}Wall-Clock\\Turnaround Time$^{**}$\end{tabular} &
  Allocated Memory &
  1\,ms &
  \begin{tabular}[c]{@{}c@{}}Memory 1\,MB\\ (CPU proportionally allocated)\end{tabular} \\ \hline
\begin{tabular}[c]{@{}c@{}}Google Cloud Run\\(Request-Based Billing)~\cite{web:gcp, web:gcp-pricing, web:gcp-instance-pricing}\end{tabular} &
  \begin{tabular}[c]{@{}c@{}}Wall-Clock\\Turnaround Time\end{tabular} &
  Allocated Memory and CPU &
  100\,ms &
  \begin{tabular}[c]{@{}c@{}}Memory 1\,MB\\ CPU 0.01\,vCPUs (1st Gen)/1\,vCPU (2nd Gen)\end{tabular} \\ \hline

\begin{tabular}[c]{@{}c@{}}Google Cloud Run\\(Instance-Based Billing)$^{***}$~\cite{web:gcp, web:gcp-pricing, web:gcp-instance-pricing}\end{tabular} &
  \begin{tabular}[c]{@{}c@{}}Wall-Clock\\Instance Time\end{tabular} &
  Allocated Memory and CPU &
  100\,ms &
  \begin{tabular}[c]{@{}c@{}}Memory 1\,MB\\ CPU 1\,vCPU\end{tabular} \\ \hline

\begin{tabular}[c]{@{}c@{}}Azure Functions Consumption Plan\\ ~\cite{web:azure, web:azure-consumption-billing, web:azure-scaling}\end{tabular} &
  \begin{tabular}[c]{@{}c@{}}Wall-Clock\\Execution Time\end{tabular} &
  Consumed Memory &
  \begin{tabular}[c]{@{}c@{}}1\,ms (min cutoff 100\,ms)\\ 128\,MB\end{tabular} &
  \begin{tabular}[c]{@{}c@{}}N/A\\ (Fixed resource size of \\ 1.5\,GB memory and 1 vCPU)\end{tabular} \\ \hline
  
\begin{tabular}[c]{@{}c@{}}Azure Functions Premium Plan$^{***}$\\ ~\cite{web:azure, web:azure-consumption-billing, web:azure-premium-billing}\end{tabular} &
  \begin{tabular}[c]{@{}c@{}}Wall-Clock\\Instance Time\end{tabular} &
  Allocated Memory and CPU &
  \begin{tabular}[c]{@{}c@{}}1\,month\\ (minimum monthly cost applies)\end{tabular} &
  \begin{tabular}[c]{@{}c@{}}CPU and Memory\\ (Fixed Combos)\end{tabular} \\ \hline

\begin{tabular}[c]{@{}c@{}}Azure Functions Flex Consumption Plan \\ ~\cite{web:azure, web:azure-flex-billing, web:azure-scaling}\end{tabular} &
  \begin{tabular}[c]{@{}c@{}}Wall-Clock\\Execution Time\end{tabular} &
  Allocated Memory &
  100\,ms (min cutoff 1\,s) &
  \begin{tabular}[c]{@{}c@{}}Memory (Either 2\,GB or 4\,GB)\\ (CPU proportionally allocated)\end{tabular} \\ \hline
\begin{tabular}[c]{@{}c@{}}IBM Cloud Code Engine Function\\ ~\cite{web:ibm, web:ibm-code-engine-billing}\end{tabular} &
  \begin{tabular}[c]{@{}c@{}}Wall-Clock\\Turnaround Time\end{tabular} &
  Allocated Memory and CPU &
  100\,ms &
  \begin{tabular}[c]{@{}c@{}}Memory (Fixed Combos)\\ CPU (Fixed Combos)\end{tabular} \\ \hline
Huawei Cloud Function Graph~\cite{web:huawei-functiongraph} &
  \begin{tabular}[c]{@{}c@{}}Wall-Clock\\Execution Time\end{tabular} &
  Allocated Memory &
  1\,ms &
  Memory (Fixed CPU-Memory Combos) \\ \hline
\begin{tabular}[c]{@{}c@{}}Alibaba Cloud Function Compute\\ ~\cite{web:alibaba-function-compute, web:alibaba-function-compute-billing, web:alibaba-usage-modes}\end{tabular} &
  \begin{tabular}[c]{@{}c@{}}Wall-Clock\\Execution Time\end{tabular} &
  Allocated Memory and CPU &
  1\,ms &
  \begin{tabular}[c]{@{}c@{}}Memory 64\,MB\\ CPU 0.05\,vCPUs\end{tabular} \\ \hline
\begin{tabular}[c]{@{}c@{}}Oracle Cloud Functions \\ ~\cite{web:oracle}\end{tabular} &
  \begin{tabular}[c]{@{}c@{}}Wall-Clock\\Execution Time\end{tabular} &
  Allocated Memory &
  Not Documented Publicly  &
  Memory (Fixed Combos) \\ \hline
Vercel Functions~\cite{web:vecel} &
  \begin{tabular}[c]{@{}c@{}}Wall-Clock\\Execution Time\end{tabular} &
  Allocated Memory &
  Not Documented Publicly &
  \begin{tabular}[c]{@{}c@{}}Memory 1\,MB\\ (CPU proportionally allocated)\end{tabular} \\ \hline
Cloudflare Workers~\cite{web:cloudflare} &
  \begin{tabular}[c]{@{}c@{}}Consumed\\CPU Time\end{tabular} &
  Consumed CPU &
  1\,ms &
  \begin{tabular}[c]{@{}c@{}}N/A\\ (Fixed resource size of 128\,MB memory)\end{tabular} \\ \hline
\end{tabular}%
}
\begin{justify}
\begin{footnotesize}
$^*$This table and related analysis in \S\ref{sec:billing-practices} focus on the most basic billable computing resources (i.e., CPU and memory). Other billable resources (e.g., storage, GPUs, and network bandwidths) may apply in practice. $^{**}$AWS bills wall-clock turnaround time that includes initialization duration starting August 2025~\cite{aws-lambda-billing-on-init}. $^{***}$Instance-based billing applies, where platforms charge for resource allocation over the function runtime instance lifespan regardless of requests.
\end{footnotesize}
\end{justify}
\caption{
\textbf{The billable models of major public serverless platforms.} The notion of billable time, billable resources, and billing granularity varies across different serverless platforms (as of 2025-05-15).}
\vspace{-2em}
\label{tab:severless-platforms-billing}
\end{table*}

To uncover these costs, a detailed analysis of current billing models together with measurements of the underlying serverless systems is required.
Previous studies have characterized major serverless platforms in terms of architecture, performance, and resource management~\cite{adzic2017serverless,wang2018peeking,yu2020characterizing,cvetkovic2023understanding}.
However, as serverless computing evolves rapidly and more serverless offerings become available, some earlier measurements do not reflect or fully capture the latest billing scheme and operation patterns (e.g., serving architecture and keep-alive behaviors) of the public serverless computing platforms.
In this work, we revisit some of the previous measurements and extend some of their performance and overhead characterization to fit modern serverless systems.

We adopt a top-down approach to analyze serverless costs.
We start with user-facing billing models and conduct large-scale trace analysis on the billing scheme.
Then, we analyze the performance, overhead, and resource allocation patterns of modern serverless request serving architectures.
Finally, we investigate the impact of OS scheduling in detail.
By tracing sources of costs from the billing model down to kernel scheduling, we provide the first comprehensive decomposition of serverless overhead and reveal the rationale behind current billing practices.
For example, our large-scale, trace-based billing model analysis reveals significant bill inflation due to wall-clock allocation-based billing (\S\ref{subsec:users-pay-more}), turnaround time billing (\S\ref{sec:turnaround-time-billing}), rounding up of resource usage and execution duration, coarse billing granularity, and high invocation fees (\Cref{tab:severless-platforms-billing} and \S\ref{sec:high-invoke-fee-and-rounding}).
Also, we investigate the dual penalty of slowdowns and higher bills stemming from the multi-concurrency model (\S\ref{sec:serving-model}), high overheads of the HTTP-based request serving architecture (\S\ref{sec:serving-arch-overhead}), and details of resource allocation during keep-alive (\S\ref{sec:ka-duration}).
Furthermore, we reveal the widespread CPU overallocation issue on public serverless platforms for the first time (\S\ref{sec:overallocation}).
Specifically, our main contributions include:

\begin{itemize}[leftmargin=*]
\item We conduct a detailed analysis on the billing practices of current major serverless platforms (\S\ref{sec:billing-practices}).
\item We analyze and quantify the overhead of modern serverless systems from several new aspects, including the concurrency model, request serving architectures, and resource allocation behaviors during keep-alive (\S\ref{sec:architecture-cost}).
\item We characterize and reveal the impact of OS scheduling granularity on major public serverless platforms (\S\ref{sec:overallocation}). 
\item We demystify the serverless billing practice through these new characterization results and analyses, and discuss implications (labeled with \textit{\textbf{I}}) for designing future performant and cost-efficient serverless systems.
\end{itemize}

We have made our artifact publicly available\footnote{\url{https://doi.org/10.5281/zenodo.17162822}}.
\section{Serverless Billing Models and Practices}
\label{sec:billing-practices}
Pay-per-use is the common billing practice on serverless platforms.
Billing models are the most direct determinants of the serverless cost as they convert billable resources (i.e., computing resources that are being billed for cloud users) into monetary charges that users immediately perceive.
Billing models vary across platforms.
In this section, we systematically deconstruct these billing practices to reveal how they shape the cost of serverless, reveal the underlying reasons for relevant billing practices, and discuss implications.

\subsection{Overview of Serverless Billing Models}
\label{sec:serverless-billing-models-overview}
\Cref{tab:severless-platforms-billing} summarizes the pay-per-use billing model on major serverless platforms listed in recent market reports~\cite{web:datadog23, web:serverless-market-report-1}.
While definitions of billable resources, wall-clock time, and pricing vary across different serverless platforms, most public serverless platforms bill a function invocation based on four factors: (1) billable wall-clock duration, (2) resource allocation amount and/or actual resource consumption over billable duration, (3) billing granularity and/or minimum billing cutoffs, and (4) a fixed fee associated with each invocation, which can be generally modeled as:
\begin{equation}
\label{eq:current-billing}
\begin{split}
\mathrm{Cost} ={}& 
\sum_{r \in R_{\mathrm{ALLOC}}}
  \left\lceil \frac{ALLOC(r)}{G_r} \right\rceil
  \times G_r
  \times \left\lceil \frac{T}{G_T} \right\rceil
  \times G_T
  \times C_r \\[6pt]
&{}+  
\sum_{r \in R_{\mathrm{USG}}}
  \left\lceil \frac{USG(r)}{G_r} \right\rceil
  \times G_r
  \times C_r
  \;+\; C_0
\end{split}
\end{equation}
where $T$ is the billable wall-clock time (e.g., wall-clock execution duration, turnaround time including initialization duration, or function instance lifespan), $R_{ALLOC}$ is the set of billable computing resources that follow allocation-based billing (e.g., vCPUs, memory, GPU, and storage), $ALLOC(r)$ defines the allocation amount of billable resources $r$ over $T$, $R_{USG}$ is the set of billable resources to which consumption-based billing applies (e.g., network bandwidths and consumed CPU time of Cloudflare Workers), $USG(r)$ defines the absolute usage amount of billable resources $r$ over $T$, $G_r$ and $G_T$ define the billing granularity of resource $r$ and wall-clock time $T$ for rounding up or minimum billing cutoff (e.g., 128\,MB and 100\,ms), $C_r$ is the per-unit price of resource $r$, and $C_0$ is the fixed invocation fee.

Depending on whether to use the whole function instance lifespan as the billable time, the billing model can generally be categorized into request-based billing (e.g., platforms other than the two listed with instance-based billing in \Cref{tab:severless-platforms-billing}) and instance-based billing (e.g., Azure Functions Premium Plan and Google Cloud Run with instance-based billing in \Cref{tab:severless-platforms-billing}). 
In request-based billing, each request is charged separately based on its execution duration (or turnaround time) and/or allocated/consumed resources during the billable period, while instance-based billing usually charges for provisioned resources on always-ready/scaled-out instances (resource allocation over instance lifespan) regardless of requests.
On most platforms, users can enable instance-based billing by changing the billing setting or configuring provisioned concurrency, minimum instances, or scale-down delay for their functions~\cite{web:gcp-instance-pricing, web:aws-reserved-concurrency}.
The fixed invocation fee ($C_0$) is usually not applied under instance-based billing.

\begin{figure}[!t]
  \centering
  \includegraphics[width=1\linewidth]{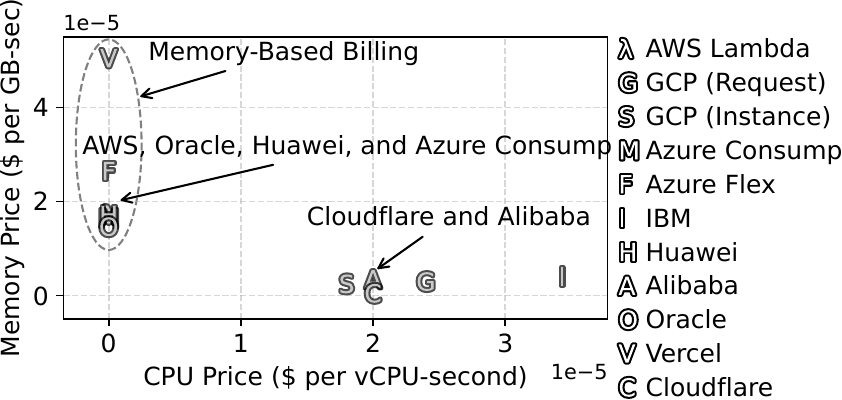}
  \caption{\textbf{Resource (i.e., vCPU and memory) prices on major serverless platforms discussed in \Cref{tab:severless-platforms-billing}.} The per-unit vCPUs and memory prices are generally similar across serverless platforms (as of 2025-05-15).}
  \label{fig:resource-prices}
\end{figure}

\Cref{fig:resource-prices} illustrates the CPU and memory prices on major serverless platforms presented in \Cref{tab:severless-platforms-billing}, which shows that the per-unit resource prices are often very similar across platforms.
Following the serverless versus non-serverless cost comparison discussed in \S\ref{sec:introduction}, this consistency in high per-unit resource prices indicates that \textbf{(\textit{I1}) the high price of serverless computing is not the result of any single provider's billing strategy} (AWS already offers some of the lowest per-unit resource prices).

\subsection{Coupled Control Knobs and Billable Resources}
\label{sec:billable-resources}
\textbf{\textit{(I2)} Control knobs and billable resources are tied closely, but allocated resources are always billed directly or indirectly:} The billable resources in serverless billing models are usually mainly defined by the available control knobs.
Public serverless platforms usually bill the computing resources they expose to users as control knobs.
For example, AWS Lambda, Vercel Functions, and the Azure Functions Flexible Consumption allocate vCPUs in proportion to the allocated memory size. 
Some platforms, such as Huawei Function Compute, Azure Functions Consumption, and Oracle Functions, offer only a fixed set of memory sizes or fixed vCPU–memory pairs, rather than fine-grained configurations (e.g., per-MB memory configuration).
In these cases, billing often appears to be based solely on memory allocation/usage, but the cost of CPU is embedded implicitly within the memory price. 
For instance, an AWS Lambda function with 1,769\,MB of memory (which corresponds to 1\,vCPU~\cite{web:aws-cpu-mem}) incurs a charge of \$$2.8792\times10^{-5}$ per second, while a GCP function (first generation with request-based billing) provisioned with 1\,vCPU and 1,769\,MB of memory costs \$$2.8319\times10^{-5}$ per second.
The price per GB-second of memory on the platforms that mainly expose and bill memory control knobs usually closely matches what one would pay for memory and CPU on platforms that allow separate CPU allocations.
Also, the ratio between the unit prices of CPU (in vCPU-seconds) and memory (in GB-seconds) on GCP, AWS Fargate (a container hosting platform that bills CPU and memory separately)~\cite{web:fargate-pricing}, and IBM Cloud Code Engine (function workloads) lies in a narrow range between 9 and 9.64, indicating a broad industry consensus on the relative value of vCPU versus memory.

On serverless platforms where vCPU and memory settings are relatively decoupled, CPU and memory are typically billed as two separate resources.
However, even these platforms impose limits on how finely resources can be tuned. 
For example, Alibaba Cloud requires the ratio of vCPU to memory (in GB) to remain between 1:1 and 1:4, with step sizes of 0.05 vCPUs and 64\,MB of memory~\cite{web:alibaba-usage-modes}.
Similarly, GCP imposes a minimum CPU allocation on the configured memory size (e.g., allocated memory of 512\,MB must be configured with at least 0.333\,vCPUs)~\cite{web:gcp-memory-limits}.
These constraints on resource control knobs usually reflect an underlying function placement challenge: highly unbalanced CPU-to-memory combinations can fragment the resource capacity on host servers, potentially leading to higher deployment costs; e.g., through decreased deployment density~\cite{cao2025making,li2023eigen}, or higher scheduling delay waiting for placement~\cite{joosen2025serverless,psychas2018randomized}.

\subsection{Inflation of Billable Resources}
\begin{figure}[!t]
  \centering
  \includegraphics[width=1\linewidth]{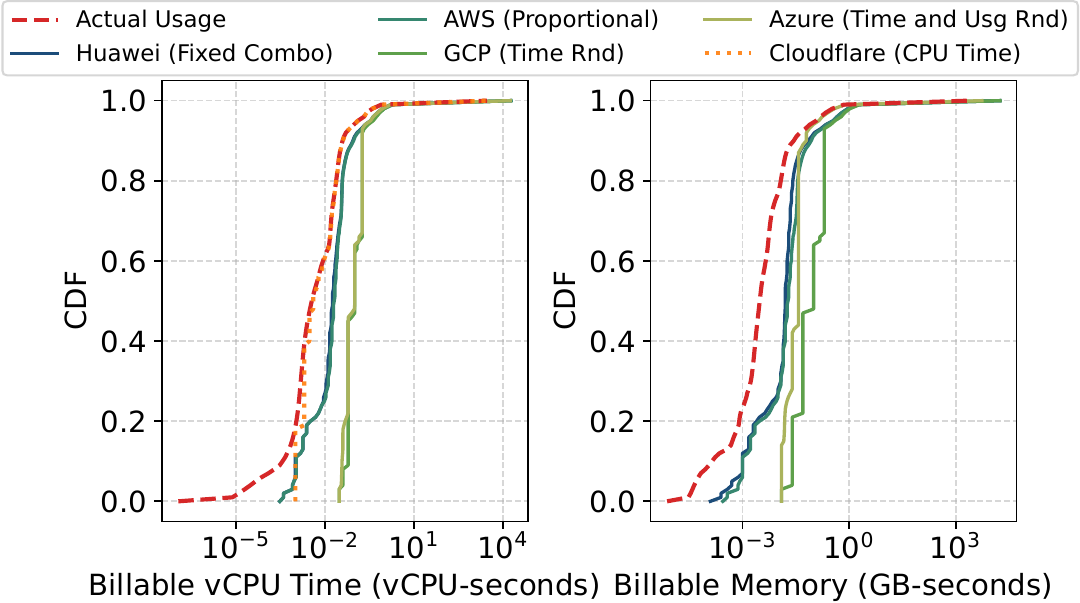}
  \caption{\textbf{Billable resources under different billing models.} The billable resources can be multiple times higher than actual consumption on major serverless platforms.}
  \label{fig:billables-current-billing}
\end{figure}

\textbf{\textit{(I3)} Billable resources are greatly inflated under wall-clock time allocation-based billing:} Understanding the amount of billable resources under different billing models is critical to evaluating the cost of serverless platforms.
To measure how much billable resources users pay on public serverless platforms, we analyze 558.74 million\footnote{The Huawei trace contains over 947.97\,million requests. We exclude the requests reporting zero CPU usage or missing valid pod IDs/flavors.} requests from the Huawei serverless trace (Huawei Public request tables)~\cite{web:huawei-cloud-data-release, joosen2025serverless} and compute the billable vCPU time and the billable memory resources of each request under the billing models presented in \Cref{tab:severless-platforms-billing}.
To avoid distortions from differences in per-unit prices on different platforms (however, they are mostly similar as discussed in \S\ref{sec:serverless-billing-models-overview}), we report raw billable resources rather than cost in dollars.
\Cref{fig:billables-current-billing} shows the distribution of these billable vCPU times and memories across requests under several representative billing models and resource allocation patterns, including proportional vCPU allocation (AWS Lambda), fixed vCPU-memory combinations (Huawei), wall-clock duration and resource usage rounding (GCP and Azure), and usage-based billing (Cloudflare Workers).
As discussed in \S\ref{sec:billable-resources}, CPU pricing is usually embedded for platforms with memory-based billing. Therefore, we include billable vCPU time for AWS.

\label{subsec:users-pay-more}

The gap between billed and actual resource usage quantifies the degree of inflated billable resources on current serverless platforms.
Our analysis reveals that, under current models, billable vCPU time exceeds actual CPU usage by a factor of $1.01\times$ (Cloudflare) up to $3.63\times$ (GCP) on average, and billable memory exceeds real memory use by $1.57\times$ (Azure) up to $4.35\times$ (GCP) on average, among which usage-based billing (Cloudflare billable CPU and Azure billable memory) shows the lowest inflation.
While differences in unit pricing shift the curve horizontally, these ratios remain the same (we do not compare absolute costs across platforms). 
Also, when mapping Huawei’s reported vCPU and memory allocations to AWS, we choose the larger of the two values to match its proportional vCPU allocation, which makes AWS billable resources slightly higher than Huawei.

\begin{figure}[!t]
  \centering
  \includegraphics[width=1\linewidth]{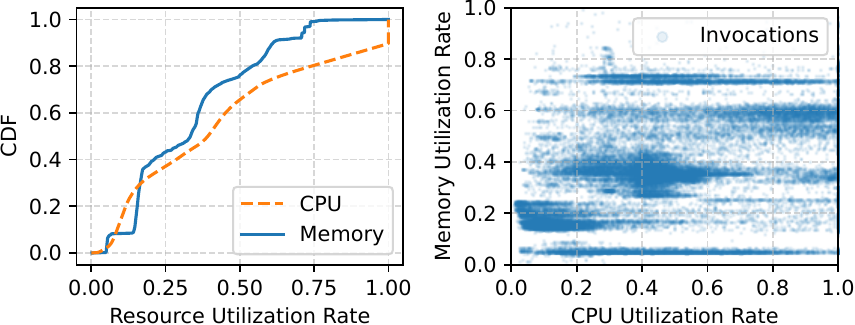}
  \caption{\textbf{Resource utilization rate distributions and their correlations.} Huawei serverless traces~\cite{web:huawei-cloud-data-release, joosen2025serverless} show that serverless functions usually have low utilization of allocated resources. Billable resources are further inflated by inflexible resource control knobs as no strong linear relationships between CPU and memory utilization rates exist.}
  \label{fig:relative-usage}
\end{figure}

One of the major driving factors of inflated billable resources is allocation-based wall-clock time billing. 
Even AWS, with one of the finest billing granularities (i.e., 1\,ms), charges billable vCPUs and memory, $2.49\times$ and $2.72\times$ higher than actual consumption on average.
Functions rarely consume their full resource allocation~\cite{joosen2023does, mahgoub2022wisefuse}, and wall-clock time includes periods when functions hold resources but remain idle or use little (e.g., blocking on remote API calls). 
\Cref{fig:relative-usage} illustrates resource usage relative to allocations. 
More than 65\% of requests use less than 50\% of the allotted CPU, and around 76\% of requests use less than half of the allotted memory.
The scatter plot of CPU and memory utilization shows a Pearson correlation of $0.552$ and a Spearman correlation of 0.565, which is slightly smaller than the value reported on serverless traces of Huawei private cloud (i.e., $0.6$) in 2023~\cite{joosen2023does}.
The moderate Pearson correlation suggests that the linear relationship is not strong, indicating that decoupling vCPU and memory configurations is crucial for reducing inflated costs.
Inflexible allocations (e.g., proportional/linear) force developers to over-provision one resource to satisfy another bottleneck~\cite{Bilal_eurosys23, verma2015large}.

Although usage-based billing offers the lowest inflation in billable resources, it currently faces limitations and has not gained widespread adoption across providers.
Cloudflare Workers is the only platform we studied that bills only on actual CPU time and aligns well with the actual CPU usage.
However, it caps code artifacts at 10\,MB and memory at 128\,MB.
This is mainly designed for small, short, single-threaded JavaScript or WebAssembly (Wasm) tasks (under 1–2\,ms) running on the V8 engine within their content delivery network (CDN)~\cite{web:cloudflare-webassembly, web:cloudflare-v8, web:cloudflare-limits}, rather than general serverless workloads.
Note that we do not argue that billing solely on the absolute amount of consumed resources is the only valid model in serverless computing.
This is because resource allocation tied to wall-clock time, particularly for resources like memory (which is risky to overcommit~\cite{tian2022owl}), significantly impacts function scheduling and deployment density.
An ideal pay-per-use billing model is one that tracks real usage, exhibiting a perfect positive correlation statistically.
There have been recent studies that aimed to move towards this ideal direction through dynamic or cooperative scheduling, or new billing models~\cite{cao2025making, zhao2024serverless, pei2024litmus, liu2023demystifying}.

\subsection{Turnaround Time Billing and Cold Starts}
\label{sec:turnaround-time-billing}
\textbf{\textit{(I4)} Billing on wall-clock turnaround time has become a common practice to compensate for the initialization phase:}
The serverless runtime sandbox lifecycle typically consists of initialization (cold start), request execution, keep-alive, and shutdown (e.g., \texttt{SIGTERM} handling)~\cite{web:aws-lifecycle, lin2024bridging}.
Depending on whether the initialization duration is included, serverless providers usually define billable wall-clock time as either execution time or turnaround time (i.e., execution time plus initialization) in their request-based, pay-per-use billing models.
Besides billing based on execution time and turnaround time, users may customize provisioned concurrency, minimum instances, or scale-down delay, and pay for the whole runtime instance lifespan (i.e., instance time) on most platforms.
Such instance time billing can further increase billable resources under bursty traffic patterns since scale-down-to-zero is delayed or disabled, and instance idle time is billed.
We observe that billing based on turnaround time has become increasingly popular on major platforms:
GCP and IBM explicitly state that they charge for the turnaround time~\cite{web:ibm-code-engine-billing, web:gcp-pricing}, while AWS recently updated its billing model to include the initialization phase (cold start delay) starting August 2025~\cite{aws-lambda-billing-on-init}.

\begin{figure}[!t]
  \centering
  \includegraphics[width=1\linewidth]{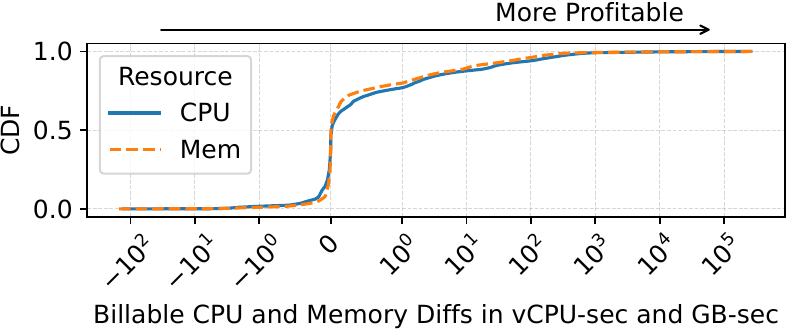}
  \caption{The differences between the billable CPU and memory resources consumed during request execution and those during initialization.}
  \label{fig:cold-start-cost}
\end{figure}

Our analysis of cold start resource usage helps explain why providers favor turnaround time billing.
We analyze 388,955 traceable cold starts from the Huawei serverless traces~\cite{web:huawei-cloud-data-release, joosen2025serverless}.
For each cold start, we consider the duration spent on the initialization of the runtime sandbox and the resource allocation.
We compute the difference between the billable resources (measured in wall-clock resource allocations) consumed during cold start and the sum of billable resources used by all subsequent requests within the sandbox.
A negative difference means the cold start alone consumed more billable resources than all later requests combined.
\Cref{fig:cold-start-cost} quantifies, for the first time, such relative resource cost of cold starts compared to subsequent executions on production systems, which shows that 42.1\% of cold starts produce a zero or negative difference.
In other words, under a billing model based on purely execution duration of requests, providers would charge less (or the same) for request execution than the actual cost of the initialization phase in about 42.1\% of cold start cases.

To avoid this revenue gap, it is a natural choice for providers to include initialization delay in the billed duration, i.e., to bill on turnaround time, which also captures variation in initialization delays (and wall-clock-based resource usage) across functions with different language runtimes and dependency requirements.
Also, providers may impose additional billing components to offset cold start costs, such as a fixed per-invocation fee and minimum billing cutoffs for billable time and resources, which we further discuss in \S\ref{sec:high-invoke-fee-and-rounding}.
Additionally, the results also show that a small portion of functions exhibit a long tail of negative resource differences, indicating that turnaround-time billing can substantially increase the cost of these functions.

\subsection{High Invocation Fee and Expensive Rounding Up}
\label{sec:high-invoke-fee-and-rounding}
\begin{figure}[!t]
  \centering
  \includegraphics[width=1\linewidth]{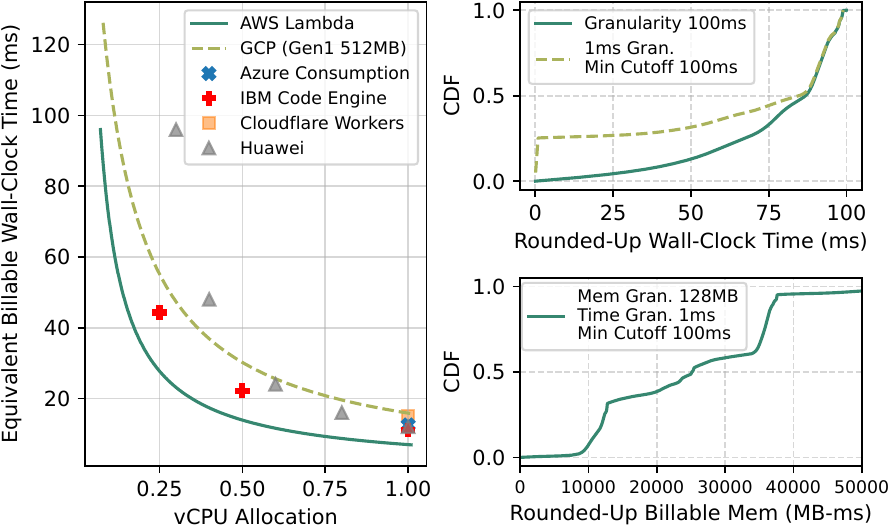}
  \caption{The equivalent billable wall-clock time of invocation fees (left) (as of 2025-05-15) and the rounded-up amount of billable wall-clock time and memory (right).}
  \label{fig:rounding-up}
\end{figure}
Major serverless platforms charge a fixed fee per invocation, typically between \$$1.5\times10^{-7}$ and \$$6\times10^{-7}$ per request~\cite{web:vecel, web:alibaba-function-compute-billing, web:lambda-pricing}.
Although these amounts seem small, they can add up disproportionately when functions run for very short durations or use minimal resources.
For example, on AWS Lambda, a fixed invocation fee of \$$2\times10^{-7}$ is equivalent to 96\,ms of billable wall-clock time for a function with the default 128\,MB memory configuration, which exceeds the average execution durations reported in the Huawei traces (i.e., 58.19\,ms)~\cite{joosen2025serverless}.
\Cref{fig:rounding-up}-left shows how invocation fees convert to equivalent billable wall-clock time across different platforms.
Besides invocation fees, several platforms apply coarse billing granularity (minimum billing increments) or cutoffs.

The charts on the right in \Cref{fig:rounding-up} present the inflated billable time and memory usage under different billing granularities for $527.05$ million requests with execution times of at least 1\,ms in Huawei traces~\cite{joosen2025serverless, web:huawei-cloud-data-release}.
For a 100\,ms billing granularity (e.g., GCP and IBM), the average rounded‑up wall-clock time is 77.12\,ms, while for a 1\,ms granularity with a 100\,ms minimum cutoff (e.g., Azure Consumption), the average rounded‑up wall-clock time is 61.35\,ms.
When billing memory with a 128\,MB granularity (e.g., Azure Consumption), the average rounded‑up billable memory is $2.67\times10^{-2}$\,GB‑seconds.
These values are on the same order of magnitude as the average execution durations and billable memory amounts reported in the studied serverless traces (58.19\,ms and $2.75\times10^{-2}$\,GB‑seconds).

Therefore, our analysis shows that \textbf{\textit{(I5)} invocation fees are high, and together with billing granularity, they can cause disproportionate costs for short, small function invocations}.
These extra costs may not be explained only as a way to offset resource usage during the initialization phase (cold start), since providers that bill for turnaround time still charge the invocation fee.
They may further be linked to overheads in the serving architecture and OS scheduling, which we analyze further in \S\ref{sec:architecture-cost} and \S\ref{sec:overallocation}.

\section{Hidden Cost of Serverless Serving Architecture}
\label{sec:architecture-cost}
Serverless computing platforms usually abstract away low-level infrastructure details. 
However, the overhead of the underlying serving layer directly affects how providers schedule and run serverless workloads and at what cost.
These are passed on to the users as the billing model and pricing parameters.
Therefore, studying serverless serving architectures is key to understanding serverless computing costs.
In this section, we analyze the serverless runtime of major serverless platforms, run benchmarks on major platforms to quantify the costs that remain hidden in the serving layer, and discuss their implications on cost.

\subsection{Cost Implications of the Concurrency Model}
\label{sec:serving-model}

Serverless platforms vary in how they map concurrent requests to sandboxes.
Depending on the maximum number of concurrent requests allowed per sandbox, there are two common serving models:
the single-concurrency model, in which the concurrency limit is strictly one (i.e., no intra-runtime concurrency), and the multi-concurrency model, where the concurrency limit can be greater than one.

In the single‐concurrency model (e.g., AWS Lambda and Cloudflare Workers), each sandbox accepts only one request at a time.
When a request arrives, the serverless platform allocates a new runtime sandbox or reuses a warm, idle sandbox for the request.
As there is no resource competition (e.g., CPU and memory) among concurrent requests, this can help keep the execution duration consistent even under high load.

Under the multi‐concurrency model, multiple requests can enter and be executed within the same sandbox concurrently, if the user code supports concurrency.
Platforms using this model usually allow users to set the maximum concurrency per sandbox and concurrency-based scaling policies.
However, \textbf{\textit{(I6)} if the extra control knob on concurrency is not configured properly, multi-concurrency can degrade function performance while increasing cost} since resource contention (e.g., CPU, memory, and cache) slows down all concurrent requests and increases billable wall-clock time (e.g., execution time) under request-based billing.
For example, running two CPU-bound requests, each requiring 1\,s of CPU time, together in a sandbox with 1\,vCPU doubles the execution duration of each request to 2\,s, thus doubling billable resources as well.
In practice, such slowdowns stemming from resource contention are often worse due to context switch overhead and cache misses~\cite{shahrad2019architectural}.

\begin{figure}[!t]
  \centering
  \includegraphics[width=1.0\linewidth]{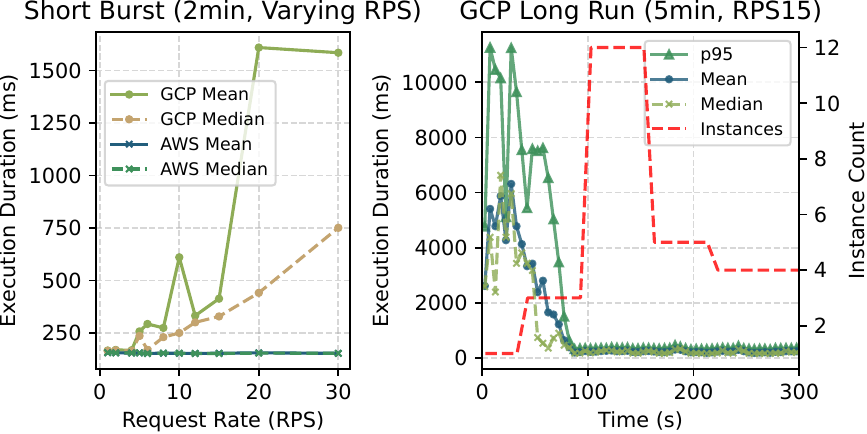}
  \caption{\textbf{Function execution durations under varying request rates.} The multi-concurrency serving model can lead to non-linear slowdowns and increased costs under high concurrency, mainly due to delays in scaling the number of sandboxes to match the incoming request load.}
  \label{fig:concurrency-model-duration-test}
\end{figure}

To show how concurrency models affect performance and cost, we deploy the same compute-intensive function (PyAES from Functionbench~\cite{kim2019functionbench}) on AWS Lambda and GCP with 1\,vCPU allocation.
Each request takes about 160\,ms of CPU time.
We use the default concurrency configurations (i.e., limit of 80) and scaling policy (60\% CPU utilization target and concurrency-based scaling~\cite{web:gcp-scaling}) on GCP.
At various request rates (in RPS), we send bursts of requests for 120\,s to simulate a short traffic spike.
The plot on the left in \Cref{fig:concurrency-model-duration-test} presents the average execution duration reported by providers.
AWS maintains a stable execution time at all request rates due to dedicated sandboxes without resource contention.
The average execution time (and cost) of the function deployed on GCP rises by up to $9.65\times$ when the request rate is higher than 6\,RPS.

Longer traffic logs reveal a key caveat of multi-concurrency models: it takes time to gather scaling metrics to scale instances to match demand.
We send a steady traffic of 15\,RPS to the GCP function for 20 minutes.
The plot on the right in \Cref{fig:concurrency-model-duration-test} shows the first five minutes (later data remains stable) of execution time and container count reported by GCP.
Scaling does not begin until about 40\,s.
This is likely due to the fact that platforms with the multi-concurrency serving model usually aggregate the scaling metrics over a time window (e.g., 60\,s by default in Knative~\cite{kpa}) to avoid oscillation~\cite{web:scaling-oscillation}.
The execution duration and instance count remain stable after around 90\,s, but the average duration is still 1.43 times higher (i.e., 239.29\,ms versus 166.78\,ms) than that under the RPS of 1 due to resource contention.

Such a dual penalty of slowdowns and higher bills stemming from the multi-concurrency model is particularly concerning, given that a recent characterization study reports that 93.3\% of serverless workloads on IBM Cloud Code Engine used Knative's default container concurrency of 100 as the limit~\cite{Nasiri_eurosys26, web:knative-concurrency}.
This suggests that most users either do not optimize, or might be unaware of concurrency settings, highlighting the critical need for relevant control knob tuning tools and concurrency optimization guidance from cloud providers.

\subsection{Request Serving Architecture Overhead}
\label{sec:serving-arch-overhead}
\begin{figure}[!t]
  \centering
  \includegraphics[width=1.0\linewidth]{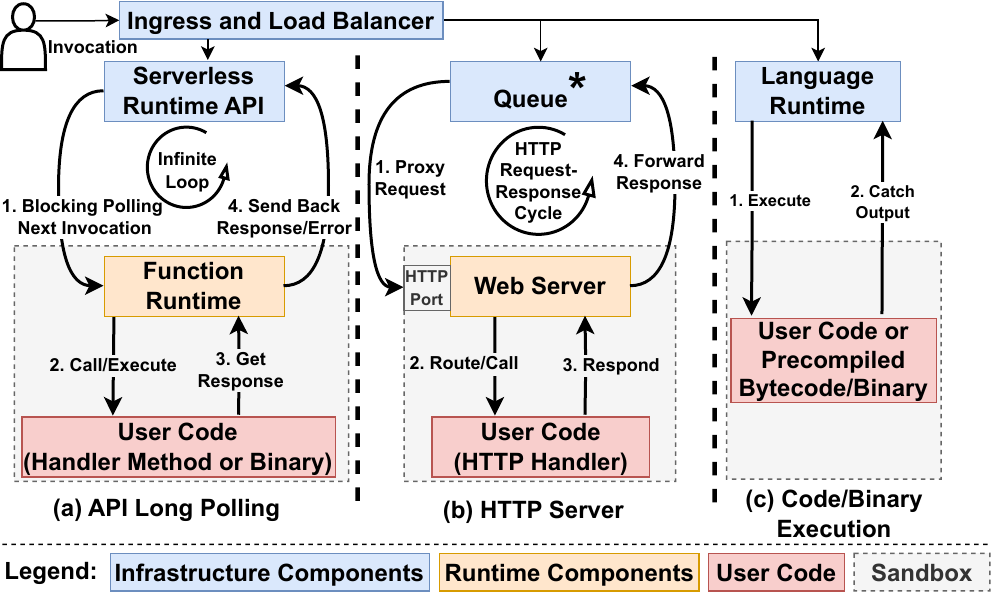}
  \begin{justify}
\footnotesize{$^*$The queue can be a runtime component as it can be located within the same scaling unit (e.g., pod) or sandbox (e.g., container) and be dedicated to each function instance on some platforms (e.g., GCP, IBM, and Knative).}
\end{justify}
  \caption{\textbf{The three mainstream serverless request serving architectures}, including (a) API long polling (e.g., AWS Lambda), (b) HTTP server (e.g., Azure and GCP), and (c) code/binary execution (e.g., Cloudflare Workers).}
  \label{fig:request-serving-architecture}
\end{figure}

\Cref{fig:request-serving-architecture} depicts three common request serving architectures on major serverless platforms.
In the runtime API long polling model (e.g., AWS Lambda~\cite{web:aws-runtime-api}), the user provides a handler method (non-HTTP) or a binary executable for processing requests. 
A runtime program (usually offered by providers) runs inside the sandbox and repeatedly polls the runtime API endpoint over HTTP or RPC in a blocking infinite loop. 
The retrieved request event is processed by the handler, and the result is posted back to the runtime API before the next poll.
In the HTTP server model (e.g., Azure, GCP, IBM, and Knative~\cite{web:azure-functions-custom-handlers, web:gcp-container-runtime-contract, web:ibm-code-engine-getting-started, web:knative}), the function itself runs an HTTP server on a given port, with the user logic wrapped in an HTTP handler.
The queue (e.g., usually running in a sidecar container) that receives the request from the ingress acts like a reverse proxy and forwards requests to the HTTP server.
In the code/binary execution architecture (e.g., Cloudflare Workers), the user uploads a code block or precompiled binary (e.g., Wasm modules~\cite{web:cloudflare-webassembly}). 
For each request, the language runtime engine (e.g., V8 JavaScript engine) compiles and executes (JIT) or loads and executes the binary, captures the output, and sends back the response.

\begin{figure}[!t]
  \centering
  \includegraphics[width=1.0\linewidth]{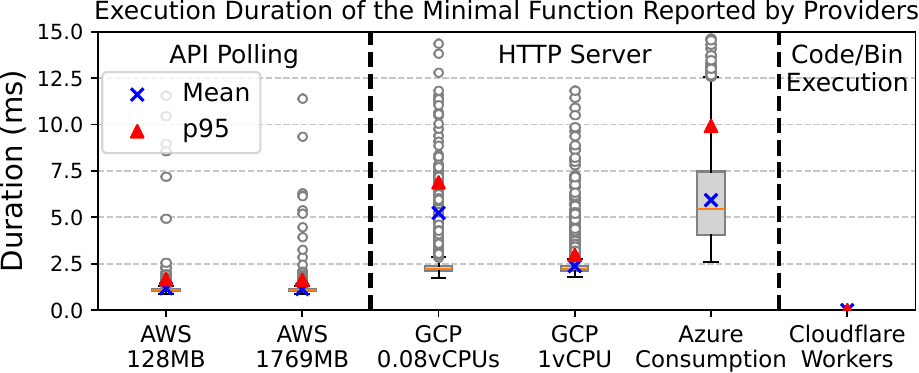}
  \caption{\textbf{Execution durations of the minimal serverless function across platforms with different serving architectures.} The functions with HTTP servers have the highest overhead, while code/binary execution has the smallest.}
  \label{fig:empty-function-latency}
\end{figure}

Each serving architecture has its own benefits.
The API polling architecture can avoid exposed ports and simplify the concurrency models by serializing event handling in each sandbox, while the HTTP server model natively supports rich HTTP semantics. 
Lastly, the code/binary execution architecture generally requires minimal runtime dependencies and artifacts.

To quantify the overhead of these architectures, we deploy a minimal serverless function that simply returns an empty string and status code across major platforms.
We measure the execution duration of the minimal function that encapsulates pod-/container-level system software.
\Cref{fig:empty-function-latency} presents the execution duration of the minimal function reported by the providers, which reflects the latency added by the request serving architecture, such as polling request events, HTTP routing, and sending back responses.
Our measurements reveal that \textbf{\textit{(I7)} platforms using the HTTP server architecture (i.e., GCP and Azure) usually have notably higher overhead, compared to API polling and code/binary execution architectures}, with an average latency up to 5.93\,ms.
This added latency can impact short functions or round the billable time up to the next interval.

This is due to that functions with the HTTP server model usually host standard HTTP servers as the upstream of the ingress, queue, and/or load balancer, which add overheads, such as maintaining HTTP listeners, connections, thread pools, and handler routing. Also, requests usually traverse additional middleware (e.g., queues, ingress, and/or service mesh sidecars) across containers and veth devices, adding latencies~\cite{cvetkovic2023understanding, zhu2023dissecting, web:gcp-knative-architecture-overview, web:gcp-deploying-container-images}.
The resource configuration may also affect such overhead (i.e., GCP 0.08\,vCPUs vs GCP 1\,vCPU), as the HTTP request–response cycle and HTTP server involve CPU-bound tasks (e.g., header and payload parsing, encoding, and serialization), and a lower CPU allocation can slow these operations.
The AWS Lambda functions that use long polling maintain a stable overhead of around 1.17\,ms on average.
Cloudflare Workers delivers near-zero latency (falling below the precision limit of 0.01\,ms reported by Cloudflare), suggesting the high efficiency of the code/binary execution architecture.

\subsection{Keep-Alive Duration and Resource Allocation Patterns}
\label{sec:ka-duration}
\begin{figure}[!t]
  \centering
  \includegraphics[width=1.0\linewidth]{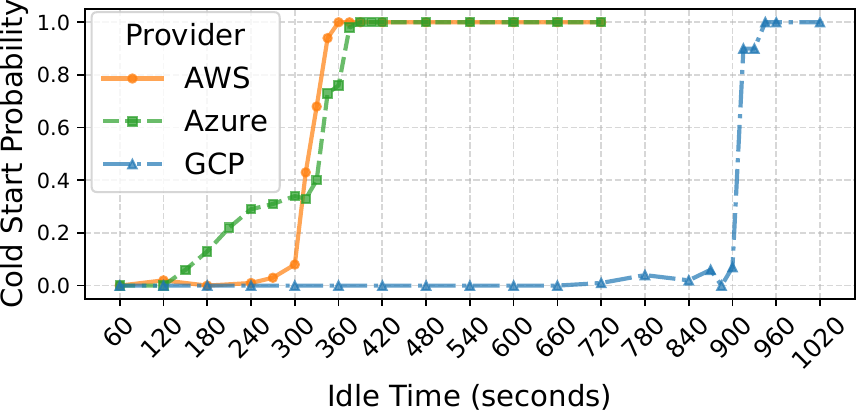}
  \caption{{\textbf{Cold start probabilities versus function idle times.}}
  The probabilities of having cold starts increase as the function sandbox idle time becomes longer. The keep-alive durations vary across platforms (as of 2025-05-15).
  }
  \label{fig:cold-start-probability}
\end{figure}

Cold start is one of the main causes of performance degradation in serverless functions, and keep-alive has become a common practice to mitigate the cold start latency~\cite{shahrad2020serverless}.
Major serverless providers keep the user function sandbox active for a period after each request to reduce the chance of cold starts for subsequent invocations.
Common keep-alive mechanisms include scale-down delay (e.g., Azure Consumption Plan, GCP, and IBM)~\cite{web:gcp-scaling}, container snapshotting~\cite{lan2024snapipeline}, code caching (e.g., Cloudflare Workers)~\cite{web:cf-cold-start-latency-mask}, and runtime freezing (e.g., AWS Lambda)~\cite{web:aws-lifecycle}.
Function keep-alive has a direct impact on provider cost, as idle functions can hold active resources (e.g., memory) or reserved capacity (for some schedulers), affecting deployment density.
Even techniques that deallocate CPU and memory during the keep-alive phase, such as snapshotting, freezing (e.g., microVM pause), and caching, require CPU time for processing snapshots and cache/storage space.
These costs are ultimately passed on to users through per-unit resource pricing or invocation fees.

We deploy serverless functions on major serverless platforms and analyze the keep-alive durations as well as the underlying keep-alive mechanisms.
We send requests at different idle intervals to check whether the sandbox is re-created and empirically measure the keep-alive duration.
The idle interval is the duration between the end of the previous invocation and the arrival of the next.
\Cref{fig:cold-start-probability} presents the probability of a cold start as a function of the sandbox idle time, calculated over 100 data points per idle interval.
The results show that AWS Lambda keeps the function sandbox alive for up to 300 to 360\,s.
Azure is likely to use an opportunistic keep-alive strategy, resulting in varying keep-alive durations between 120\,s and 360\,s.
Also, Azure pre-warms the function if the platform detects cold starts occurring at regular intervals (i.e., through idle time histograms)~\cite{shahrad2020characterization, web:azure-prewarm}.
However, we did not observe such behavior in our experiments, despite regular traffic patterns, as we encountered consistent cold starts at high idle times.
This is probably due to the test period being too short for Azure to learn traffic patterns~\cite{shahrad2020serverless}.
Besides, Azure may further increase the keep-alive duration for functions with higher traffic and that have been scaled up to multiple instances.
We observe a maximum keep-alive duration of around 740\,s for the Azure function that has scaled up to 3 instances. 
In contrast, GCP has the longest keep-alive duration, with the most instances being kept alive for about 900\,s.
Compared with the data reported in 2018 (e.g., AWS Lambda usually kept functions alive for up to 27 minutes)~\cite{wang2018peeking}, our observations reveal that \textbf{\textit{(I8)} keep-alive durations on current serverless platforms still vary but have become shorter than previous measurements}, possibly reflecting opportunistic strategies or measures for cost savings.

\setlength{\textfloatsep}{5pt}
\begin{table}[]
\resizebox{\columnwidth}{!}{%
\begin{tabular}{|c|c|c|}
\hline
\textbf{\begin{tabular}[c]{@{}c@{}}Serverless \\ Platform\end{tabular}} &
  \textbf{\begin{tabular}[c]{@{}c@{}}Keep-Alive\\ Phase Behavior\end{tabular}} &
  \textbf{\begin{tabular}[c]{@{}c@{}}Graceful Shutdown\\ after Keep-Alive\end{tabular}} \\ \hline
AWS Lambda &
  \begin{tabular}[c]{@{}c@{}}Deallocate CPU and memory\\ (Freeze and Resume)\end{tabular} &
  \begin{tabular}[c]{@{}c@{}}Supported with Lambda Extensions\\ (wait for SIGTERM handling)~\cite{web:lambda-extensions}\end{tabular} \\ \hline
\begin{tabular}[c]{@{}c@{}}GCP Function \\ (Request-Based Billing)\end{tabular} &
  \begin{tabular}[c]{@{}c@{}}Scale down CPU\\ (to about 0.01\,vCPUs)\end{tabular} &
  \begin{tabular}[c]{@{}c@{}}N/A\\ (kill without SIGTERM)\end{tabular} \\ \hline
\begin{tabular}[c]{@{}c@{}}Azure Function\\ (Consumption)\end{tabular} & Run as usual   & \begin{tabular}[c]{@{}c@{}}N/A \\ (kill right after SIGTERM)\end{tabular} \\ \hline
Cloudflare Workers                                                     & Code/Bytecode Cache & N/A                                                                        \\ \hline
\end{tabular}%
}
\caption{
The resource allocation behavior during keep-alive varies across platforms (as of 2025-05-15).
}
\label{tab:ka-phase-behavior}
\end{table}

To examine resource consumption during keep-alive, we run CPU profiling workloads (i.e., \Cref{alg:profile} discussed in \S\ref{sec:overallocation}) and empirically measure the CPU resources available to sandboxes during the keep-alive phase.
\Cref{tab:ka-phase-behavior} summarizes the resource allocation patterns of the function sandbox during keep-alive and its graceful shutdown behavior when exiting the keep-alive phase and being terminated across platforms.
\textbf{\textit{(I9)} The resource allocation behavior during keep-alive varies across platforms, and so do its performance and cost implications for serverless providers and users.}

AWS freezes the sandbox (e.g., puts the microVM to sleep) and resumes it when a new request arrives~\cite{web:aws-sleep}.
Therefore, no active CPU or memory resources are allocated during keep-alive.
On GCP, CPU allocation is dynamically scaled down to around 0.01\,vCPUs during keep-alive and is scaled back up to the user-configured level when requests arrive within the keep-alive window.
Both AWS and GCP deallocate or scale down resources during keep-alive, which naturally saves cost.
In contrast, Azure appears to make no change to CPU and memory allocation during keep-alive, which may explain its shorter, opportunistic keep-alive period that reflects a trade-off between resource use and cold start probabilities.
Cloudflare pre-warms functions on receiving the TLS handshake before the connection establishment, which can mask the very short loading and JIT compilation latency (e.g., around 5\,ms) in case of cold starts~\cite{web:cf-cold-start-latency-mask}.

The keep-alive duration and behaviors can directly affect function performance and user costs.
A longer keep-alive period can help reduce user costs on platforms that charge based on turnaround time, which includes the cold start latency.
Also, keeping resources and sandboxes active during keep-alive (e.g., Azure and GCP) can enable the execution of background tasks and increase the probability of reusing long-lived, persistent connections.
Deallocating resources (e.g., AWS and Cloudflare) may cause remote servers to close connections when they stop receiving heartbeat packets, adding overhead and cost when connections must be re-established.
Furthermore, Azure maintains full resource allocation during keep-alive, enabling user-created background tasks to run outside the request execution window.
Since Azure Consumption is billed on memory consumption during request execution, resource consumption during keep-alive is not billed (although background tasks can still affect billable wall-clock time and memory usage during request executions due to resource contention)~\cite{web:azure-backgroud-task-billing}.
This may provide opportunities for users to exploit resource allocation.
For example, a short request can start a background task running in another thread or coroutine.
The background task can send results to other cloud services (e.g., block storage) or remote endpoints after completion, or a subsequent request can retrieve those results.
We have successfully implemented this execution pattern on Azure.
By doing so, only one or two brief requests are billed, which can substantially reduce overall cloud costs.

\begin{figure*}[htb]
  \centering
  \begin{subfigure}[b]{0.49\textwidth}
    \includegraphics[width=\linewidth]{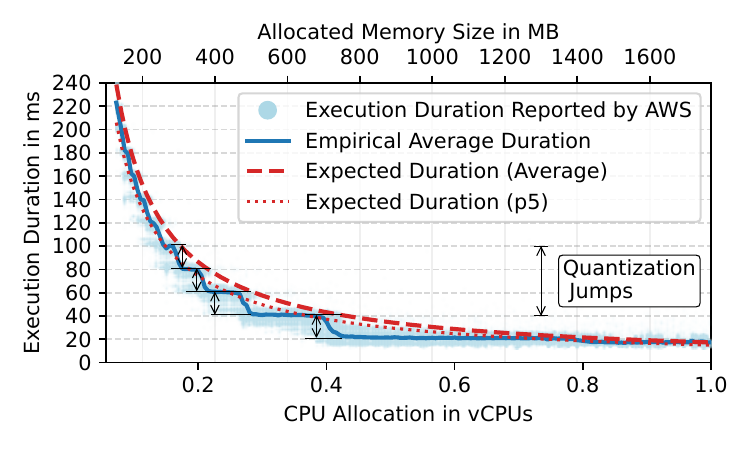}
    \vspace{-0.6cm}
    \caption{AWS Lambda}
  \end{subfigure}
  \hfill
  \begin{subfigure}[b]{0.49\textwidth}
    \includegraphics[width=\linewidth]{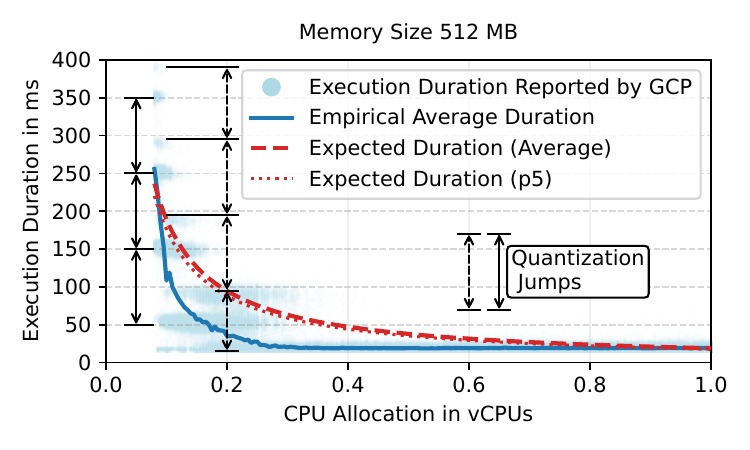}
    \vspace{-0.6cm}
    \caption{Google Cloud Run Function (GCP)}
  \end{subfigure}
  \vspace{-0.2cm}
  \caption{\textbf{Function execution durations and varying fractional CPU allocations.} The difference between the ideal (expected) and actual execution duration shows CPU overallocation for functions hosted on major serverless platforms. GCP logs show two sets of quantization jumps, which may be the cause of CPU scaling down/up when entering/exiting keep-alive phases (\S\ref{sec:ka-duration}).}
  \label{fig:pyaes-overallocation}
\end{figure*}

\section{Impact of OS Scheduling on Performance and Cost}
\label{sec:overallocation}

Serverless has a high degree of co-tenancy on servers compared to traditional VM hosting environments~\cite{shahrad2019architectural,copik2021sebs}.
In this environment, the OS kernel plays a crucial role in enforcing resource isolation and fair allocation across workloads with varying limits from different tenants.
Common approaches involve fairness-oriented schedulers (e.g., CFS and EEVDF) and control groups (cgroups) for CPU bandwidth control and resource isolation.
We observe that when the execution time of a function, the required CPU time, and the billing granularity all fall within the same range as the OS timer tick, scheduling can significantly impact performance and costs.
For the first time, we carefully characterize and understand these effects on public serverless platforms.

\subsection{Overallocation on Public Serverless Platforms}
\label{subsec:overallocation-on-public-platform}
We deploy a single-threaded, compute-bound serverless function (PyAES from Functionbench~\cite{kim2019functionbench}) on AWS Lambda under memory sizes ranging from 128\,MB (minimum size) to 1,769\,MB, and on GCP (first generation is used due to its support for fractional vCPU allocation) under CPU configurations ranging from 0.08 (minimum size) to 1\,vCPU.
AWS Lambda allocates vCPUs proportionately to the configured memory size, with 1,769\,MB equivalent to 1 vCPU~\cite{web:aws-cpu-mem,zhang2021faster}, while GCP provides a fine-grained CPU control knob with a 0.01\,vCPUs increment~\cite{web:gcp-cpu-limits}.
\Cref{fig:pyaes-overallocation} shows the execution duration reported by the serverless platform with 900,000 samples in total under different CPU configurations.
We make two main observations based on these real-world execution logs.

First, a single-threaded, CPU-bound workload like PyAES with a fractional vCPU allocation should experience a slowdown of $\frac{1}{{{\rm{vCPU Fraction }}}}$ that follows reciprocal scaling (i.e., half the core allocation, double the execution time). 
However, the empirical average (solid blue line) is consistently less than the expected average duration (dashed red line) on AWS and GCP (except for a few of the smallest vCPU allocations on GCP).
The expected average and expected 5th percentile shown in the figure are based on measurements at full vCPU allocation, scaled proportionally for smaller resource allocations.
In other words, a function can ask for, say, half the resources, but be less than twice as slow. 
In cost terms, this means that users may be charged less than expected under the current wall-clock time billing models presented in \Cref{eq:current-billing}.

Second, the average empirical execution duration does not have a smooth, reciprocal decline with increasing resource allocations.
Instead, it falls with sudden drops, which become less frequent at higher resource allocations.
These sudden drops create considerable performance jitters.
Also, this means that the allocation-based component in the billing model (i.e., $R_{ALLOC}$ in \Cref{eq:current-billing}), also known as the capacity cost, can be reduced by choosing smaller resource limits.
We observed this pattern in other functions too, and it is more pronounced with increased compute-boundness.

The performance patterns shown in \Cref{fig:pyaes-overallocation} give us clues into what might be going on. 
Reducing the resource allocation of the AWS Lambda function from 1\,vCPU at first does not affect the performance of the function, but suddenly there are increases at slightly above 1400\,MB, 700\,MB, 470\,MB, 350\,MB, 280\,MB, and so on.
These follow a scaled harmonic sequence: \textasciitilde$1400\times\{1, \frac{1}{2}, \frac{1}{3}, \frac{1}{4}, \frac{1}{5}, ...\}$.
This discrete $\frac{1}{n}$ sequence suggests the presence of a quantization effect, rather than the continuous proportional allocation ($\frac{1}{x}$) initially expected.
Namely, the function is sometimes given more than it is supposed to receive, since the underlying CPU allocation units are quantized, causing jumps on the performance curve.
As an analogy, if you want 2 kg of sugar and it is sold in 1 kg packs, the seller gives you two packs. However, if you ask for 1.5 kg, the seller would still need to give you 2 packs, leaving you with an extra 0.5 kg (i.e., overallocation).
We observe the same quantization-based overallocation on major serverless platforms.

\subsection{Quantized OS Scheduling}
\label{sec:quantization-comes-from-os}

By default, the Linux kernel leverages the Completely Fair Scheduler (CFS) or the Earliest Eligible Virtual Deadline First (EEVDF) scheduler (default scheduler since Linux kernel 6.8) to allocate resources in a fair or latency-sensitive manner~\cite{web:eevdf-intro}.
The scheduler generally provides each runnable process with a baseline allocation of resources (e.g., CPU time slice), ensuring that it receives at least one opportunity to execute on the processor.
It also incorporates mechanisms like CPU Bandwidth Control~\cite{turner2010cpu} and cgroups~\cite{web:cgroups} to impose resource limits and provide resource isolation.
Such mechanisms have become the foundation of resource isolation and allocation in the sandboxing solutions widely deployed in serverless, such as containers~\cite{web:container-runtime, li2022rund}, microVMs~\cite{agache2020firecracker}, and Wasm~\cite{shillaker2020faasm}.
Our observations in \S\ref{subsec:overallocation-on-public-platform} are the results of the existing allocation slices in the OS scheduler and cgroups, which seem to be coarse-grained for increasingly short serverless functions, causing issues with cost fairness and performance variability.

The OS maintains a kernel data structure (\texttt{cfs\_bandwidth}) for bandwidth control of each cgroup (\texttt{task\_group}), which includes information such as the enforcement period (CFS period), runtime quota (CFS quota) within each period, remaining runtime available for use (global runtime pool) protected by a spinlock, as well as the throttled run queue~\cite{web:cfsbandwidth-datastructure}.
Note that the newer Linux kernels with the EEVDF scheduler use a similar interface and kernel data structure for CPU bandwidth control as CFS.
Therefore, the CFS period and quota we discuss in this section also apply to kernels with the EEVDF scheduler.
For the rest of the section, we refer to them as (CPU bandwidth control) period and quota.

\begin{figure}[!t]
  \centering
  \includegraphics[width=1.00\linewidth]{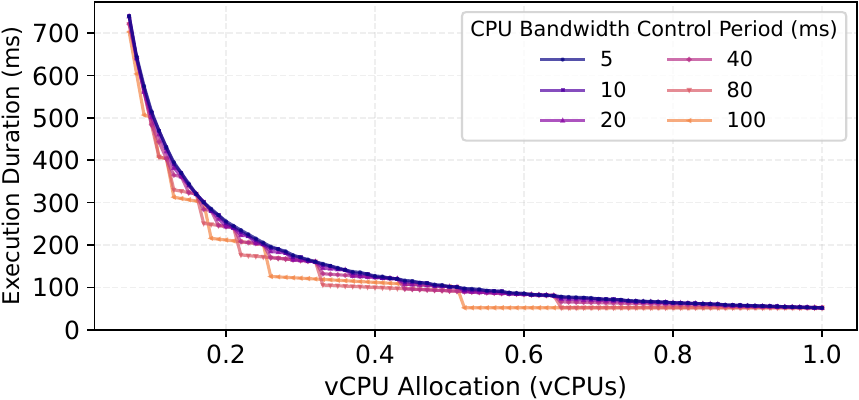}
  \caption{\textbf{Theoretical execution durations under fractional CPU allocations.} Shorter CPU bandwidth control periods improve degradation proportionality for sub-core allocations.
  }
  \label{fig:ideal-cfs-equation}
\end{figure}

A high-resolution timer (\texttt{hrtimer}) is registered with a callback~\cite{web:hrtimer-callback} to refill the global pool with the quota once per period. 
Each logical CPU core within the cgroup has a local pool of available runtime for per-CPU-basis runtime accounting. 
During runtime accounting (e.g., at scheduler ticks or context switches), the consumed runtime is subtracted from the local pool for processes running on a core within the cgroup.
When the local pool runs out of runtime, it attempts to acquire more (the smaller of \texttt{sched\_cfs\_bandwidth\_slice}~\cite{web:cfs-kernel-doc} or remaining runtime) from the global pool.
If both the global and local pools are exhausted, processes on the core are throttled and moved to the throttled run queue.
When the global pool is refilled to have available runtime in the new period (\texttt{hrtimer} callback), the scheduler distributes runtime among throttled run queues and unthrottles them (i.e., marks them as eligible to be scheduled again).
Under this schema, the wall clock duration of a CPU-bound process can be calculated as:
\begin{equation}
\label{eq:idea-cfs}
d=
\left\{
\begin{array}{ll}
\left\lfloor {{\raise0.7ex\hbox{$T$} \!\mathord{\left/
 {\vphantom {T Q}}\right.\kern-\nulldelimiterspace}
\!\lower0.7ex\hbox{$Q$}}} \right\rfloor  \times P + T\bmod Q & \mbox{if $T\bmod Q \ne0$},\\
\left( {\left\lfloor {{\raise0.7ex\hbox{$T$} \!\mathord{\left/
 {\vphantom {T Q}}\right.\kern-\nulldelimiterspace}
\!\lower0.7ex\hbox{$Q$}}} \right\rfloor  - 1} \right) \times P + Q & \mbox{otherwise}
\end{array}
\right.
\end{equation}
Here, $d$ is the execution duration, $T$ is the required CPU time, $P$ is the period, and $Q$ is the quota.
The scheduler tries to limit the CPU utilization of tasks under CPU bandwidth control to $Q/P$.
\Cref{fig:ideal-cfs-equation} shows the execution durations derived by \Cref{eq:idea-cfs} for a CPU-bound workload with a CPU time of 51.8\,ms (the average value\footnote{The requests that report zero CPU usage are excluded.} in Huawei serverless traces~\cite{joosen2025serverless,web:huawei-cloud-data-release}) under different periods from 5\,ms to 100\,ms and the quotas mapped by varying fractional vCPU allocations.
These periods are in the same scale compared to those we found empirically (shown later in \S\ref{sec:scheduling-profiling}).
With longer periods, the quantization effect becomes more pronounced.
As periods decrease, the execution duration converges to the ideal execution duration following reciprocal scaling.

The model above does not account for the fact that the runtime accounting and throttling mechanisms cannot operate with infinite frequency or precision due to excessive overhead (e.g., handling \texttt{hrtimer} interrupts~\cite{web:hrtimer-interrupt-overhead}) in real-world systems.
Since the scheduling tick frequency is usually between 100 and 1,000\,Hz (\texttt{CONFIG\_HZ})~\cite{pellegrini2015time, web:config-hz}, runtime accounting and task group throttling is often delayed, especially with the relatively long scheduler tick frequency (e.g., 250\,Hz or less).
Therefore, a task may often consume runtime more than the quota within a period (overrun) due to lagged accounting, resulting in a negative runtime in the local pool~\cite{ugedal2022mitigating}.
In this case, the task may be throttled for one or more periods to wait for the quota refill and pay back the runtime debt.
For example, consider a CPU-bound task within cgroup with 1.45\,ms quota over 20\,ms period (i.e., 0.072\,vCPU allocated to AWS Lambda with 128\,MB memory) and tick interval of 4\,ms (250\,Hz).
A possible scenario is that it first gets 4\,ms CPU time and is throttled for 36\,ms (rest of the first period and the whole second period) and becomes eligible to run again in the third period (after 40\,ms).
Then, the task runs another 4\,ms after the quota is refilled, causing overrun again with more debt, and is throttled for 56\,ms until 100\,ms and so on.
This task repeatedly alternates between running for 4\,ms and being throttled for long periods (i.e., 36\,ms or 56\,ms) over multiple periods due to overrun and lagged accounting.

Modern kernels often run with the tickless mechanism, with less frequent scheduling interrupts under light loads~\cite{web:nohz, siddha2007getting}.
Also, scheduling decisions and runtime accounting do not occur only at scheduler ticks.
Events like voluntary context switches or interrupts (e.g., \texttt{hrtimer}) can also trigger accounting, rescheduling, or preemption.
This can lead to variations in runtime allocation and throttled duration.
Overrun issues marginally impact long tasks as the OS scheduler ensures fairness over time, but can significantly affect short tasks.
However, a defining feature of serverless is the short execution for the majority of requests~\cite{shahrad2020serverless,joosen2025serverless,joosen2023does}.
Therefore, even without the aforementioned overrun effect, CPU overallocation can still happen if a serverless workload is shorter than the CPU bandwidth control enforcement period.
For example, a task that requires 10\,ms CPU time running within a cgroup with a 20\,ms period of a 10\,ms quota is allowed to consume 100\% of the CPU during its brief execution, regardless of the configured limit of 0.5\,vCPUs.
For relatively long tasks that span multiple periods, such overallocation can still happen within the last period before the task is finished.
I/O-bound tasks are usually blocked, usually not using CPU while waiting for I/O (e.g., \texttt{epoll\_wait()}).
However, when the task resumes after data becomes available, overruns and throttling across periods may occur, though this is less pronounced as the task uses the CPU intermittently, consuming less runtime and triggering fewer throttles.
In a word, \textbf{\textit{(I10)} current OS scheduling granularity seems to be coarse in the context of serverless computing}.

\begin{figure*}[!t]
  \centering
  \begin{subfigure}[b]{0.49\textwidth}
    \includegraphics[width=\linewidth]{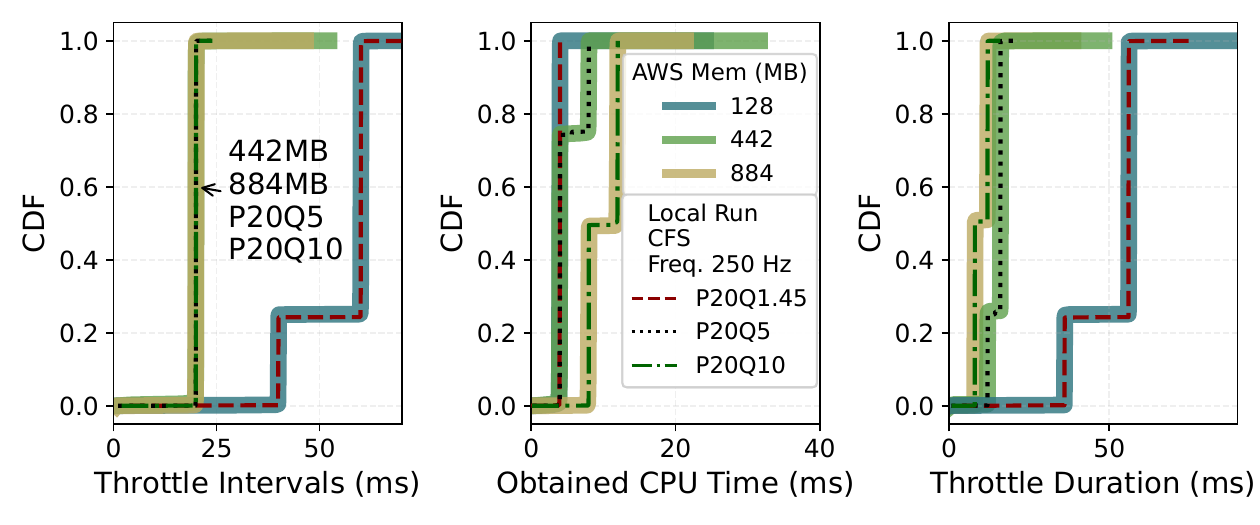}
    \caption{AWS Lambda}
  \end{subfigure}
  \hfill
  \begin{subfigure}[b]{0.49\textwidth}
    \includegraphics[width=\linewidth]{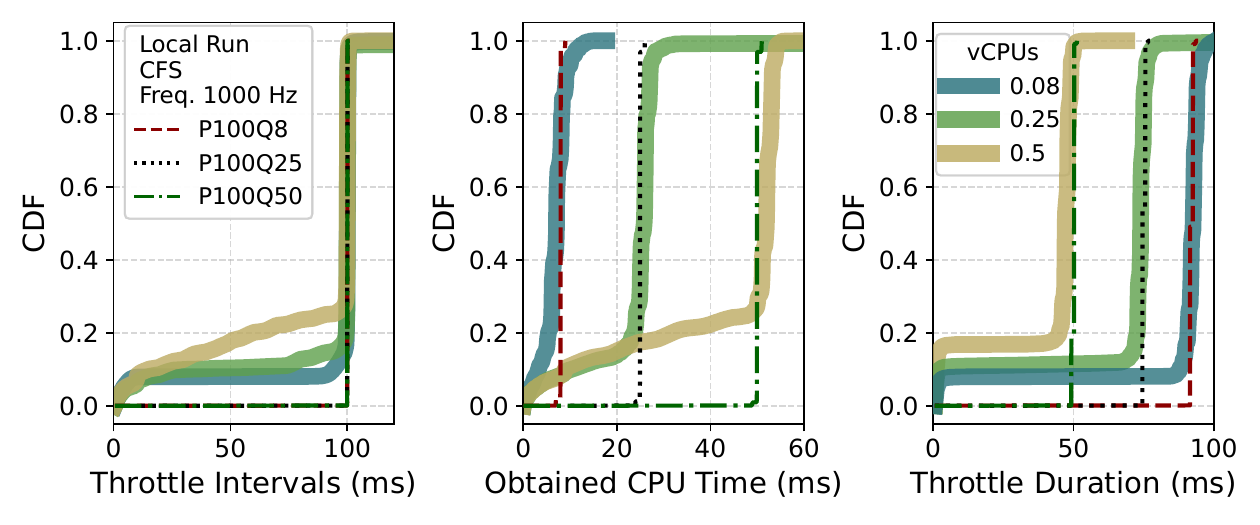}
    \caption{Google Cloud Run (GCP)}
  \end{subfigure}
  \begin{subfigure}[b]{0.49\textwidth}
    \includegraphics[width=\linewidth]{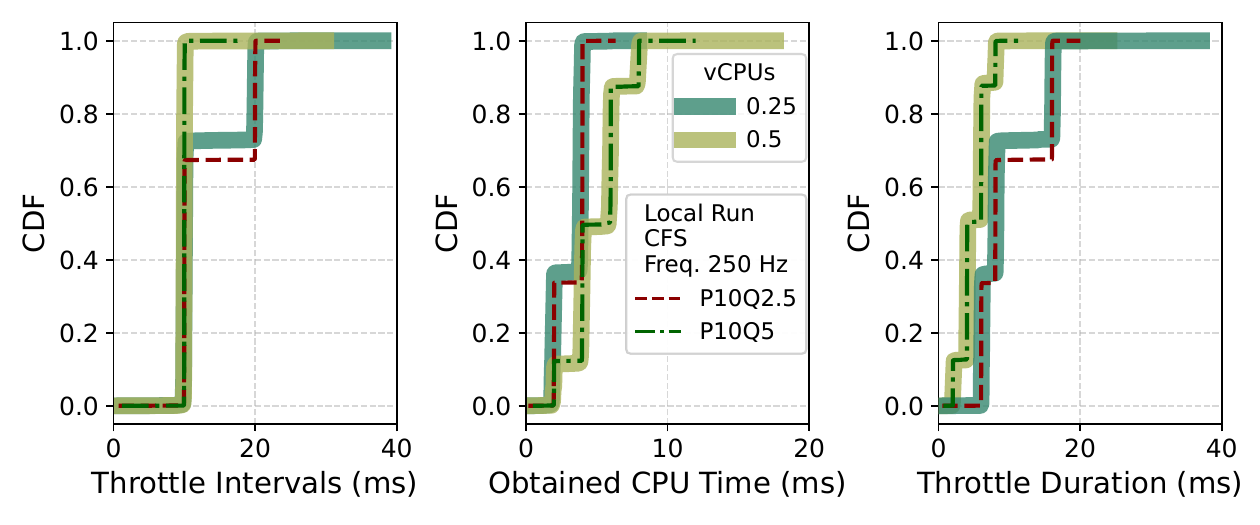}
    \caption{IBM Code Engine}
  \end{subfigure}
  \begin{subfigure}[b]{0.49\textwidth}
    \includegraphics[width=\linewidth]{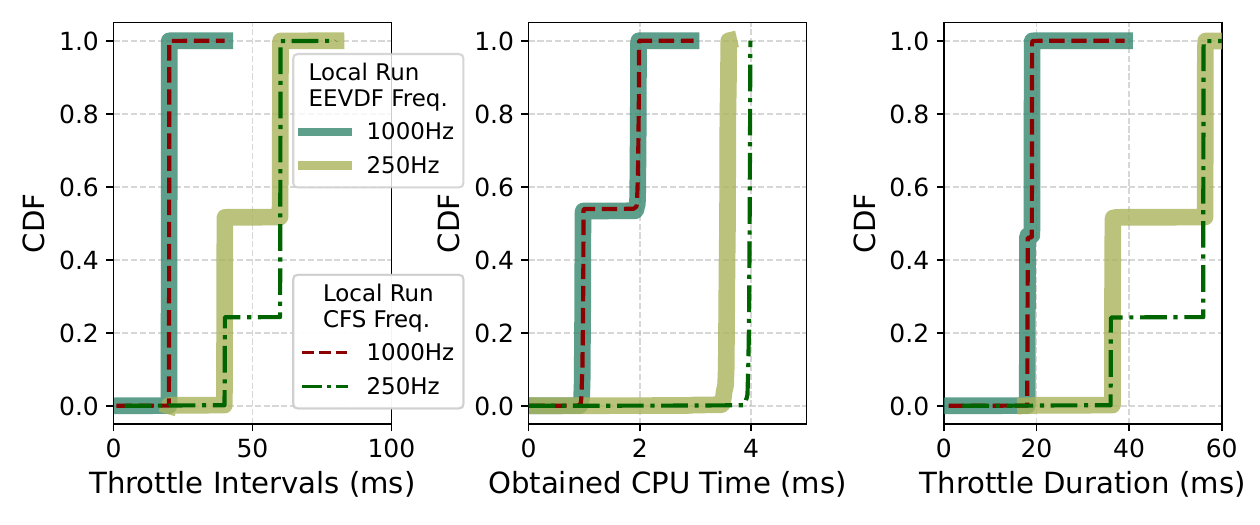}
    \caption{CFS vs EEVDF under Different Timer Frequencies (P20Q1.45)}
  \end{subfigure}
  \begin{justify}
\footnotesize{Note: The dashed and dotted lines are results of local runs with configurations that match the cloud profiling results most. The numbers following P and Q in the legend stand for CPU bandwidth control period and quota in milliseconds. The legend also shows the scheduler and the timer frequency of local runs.}
\end{justify}
\vspace{-0.2cm}
  \caption{\textbf{Distributions of throttle intervals, throttle durations, and obtained CPU times (runtime) under the studied scheduling settings.} We successfully match the local scheduling setting to cloud deployments. The scheduler profiling results (figures (a), (b), and (c)) reveal that the scheduling settings and granularity vary across serverless platforms.
  }
  \label{fig:cloud-platform-schedprofile-results}
\end{figure*}

\subsection{Scheduling Granularity of Serverless Platforms}
\label{sec:scheduling-profiling}

\begin{algorithm}[!t]
\small
\caption{Profile Runtime and Throttle}\label{alg:profile}
\begin{algorithmic}[1]
\State $s \gets get\_clock\_monotonic()$\Comment{Get monotonic clock time}
\State $n\_throt \gets 0$
\State $THRO \gets []$\Comment{Array of tuples of throttle detected time and throttle duration}
\State $last\_chkpt \gets s$
\While{$true$}
\State $now \gets get\_clock\_monotonic()$
\If{$now - last\_chkpt \geq 500$$\mu s$}
    \State $THRO[n\_throt++] \gets (now, now - last\_chkpt)$
\EndIf
\State $last\_chkpt \gets now$
\If{$now - s \geq EXEC\_DUR $}
    \Return $THRO$
\EndIf
\EndWhile
\end{algorithmic}
\end{algorithm}

The observations and discussions in \S\ref{subsec:overallocation-on-public-platform} and \S\ref{sec:quantization-comes-from-os} prompt us to further investigate the OS scheduling settings of major serverless platforms and their impact on performance and cost.
We analyze three major serverless providers, namely AWS Lambda, GCP, and IBM.
However, public serverless providers abstract away infrastructure details and do not expose the underlying scheduling mechanisms and parameters~\cite{jonas2019cloud, mampage2022holistic}.
Therefore, we run functions on target platforms to profile the scheduling behaviors and empirically peek at their scheduling behaviors from the user space.

\textbf{\textit{Methodology}}: \Cref{alg:profile} presents the pseudocode of the scheduler profile function, in which the function runs for a predefined duration (\texttt{EXEC\_DUR}) and records the time and value of sudden increases (>500\,$\mu s$) in monotonic clock time (\texttt{CLOCK\_MONOTONIC}) readings.
The default minimal preemption granularity for CPU-bound tasks in the kernel is 750\,$\mu s$~\cite{web:min-preempt-gran}, and such time jumps can effectively suggest the occurrence of throttles.
We invoke the function with different vCPU configurations, each with 300 invocations.
Each function request runs for 10\,s, leading to runtime/throttle data collected over 3,000\,s of execution span for each configuration.
Additionally, to be able to assess the effect of different quotas, periods, and OS schedulers, we use in-house VMs, each with 10 vCPUs (Intel Xeon E5-2673 v4), Linux kernel 6.2 (CFS) or 6.8 (EEVDF scheduler), and the timer frequency of either 250\,Hz or 1,000\,Hz, to profile the function within containers (\texttt{runC} runtime).
We analyze the interval between throttles, the throttle duration, and the consumed CPU time before each throttle by calculating the differences between consecutive events in the recorded data.

\textbf{\textit{Empirical Analysis}}: 
Figures~\ref{fig:cloud-platform-schedprofile-results}(a) to (c) present the distribution of throttle intervals, durations, and obtained CPU time (runtime) of the studied settings.
AWS Lambda functions have throttle intervals that are multiples of 20\,ms, whereas IBM functions show multiples of 10\,ms.
The interval, duration, and runtime results closely align with local runs with corresponding vCPU allocations, periods of 20\,ms (for AWS) and 10\,ms (for IBM), and the timer frequency of 250\,Hz.
Also, the runtime and throttle duration of the AWS function (128\,MB, 0.072\,vCPUs) and their distributions align with the theoretical analysis discussed in \S~\ref{sec:quantization-comes-from-os}.
The quantized obtained CPU time of AWS Lambda suggests a coarse scheduling granularity under a lower timer frequency (i.e., 250\,Hz).
The overrun almost happens every time the task is scheduled.
Functions on IBM show similar quantized scheduling patterns.
The GCP functions exhibit throttle intervals of 100\,ms in most cases, while they have 6.42\% - 14.83\% of throttle durations shorter than 2\,ms, indicating frequent context switches and preemption events even within the CPU bandwidth control quota.
Compared to other platforms, the less quantized obtained CPU time (i.e., a smoother curve without distinct step-like jumps as shown in ~\Cref{fig:cloud-platform-schedprofile-results}(b)-Mid) indicates finer-grained time slice allocation under a higher timer frequency.
\Cref{tab:severless-platform-profiling} presents the scheduling parameters obtained by our empirical analysis, which suggest that public cloud providers do not have a unanimous configuration.

\setlength{\textfloatsep}{5pt}
\begin{table}[!t]
\resizebox{\columnwidth}{!}{%
\begin{tabular}{|c|c|c|}
\hline
\textbf{\begin{tabular}[c]{@{}c@{}}Serverless\\ Platform\end{tabular}} & \textbf{\begin{tabular}[c]{@{}c@{}}Bandwidth Control Period\\ (\texttt{cfs.cpu\_period})\end{tabular}} & \textbf{\begin{tabular}[c]{@{}c@{}}Scheduler Tick Freq\\ (\texttt{CONFIG\_HZ})\end{tabular}} \\ \hline
AWS Lambda                                                             & 20\,ms                                                                            & 250                                                                                 \\ \hline
\begin{tabular}[c]{@{}c@{}}Google Cloud Run\\ Functions\end{tabular}   & 100\,ms                                                                           & 1000                                                                                \\ \hline
\begin{tabular}[c]{@{}c@{}}IBM Cloud Code \\ Engine Functions\end{tabular}   & 10\,ms                                                                            & 250                                                                                 \\ \hline
\end{tabular}%
}
\caption{
Scheduling parameters obtained by empirical analysis (as of 2025-05-15), which vary across different providers.
}
\label{tab:severless-platform-profiling}
\end{table}

\textbf{\textit{Does the new EEVDF scheduler solve the overallocation issue?}}
The EEVDF scheduler has replaced the CFS scheduler in Linux kernel version 6.8, which introduces a virtual deadline mechanism that improves system responsiveness by prioritizing latency-sensitive tasks with shorter time slices~\cite{stoica1995earliest, web:eevdf-intro}.
However, overrun issues still persist under EEVDF because runtime accounting and scheduling granularity remain tied to the timer frequency.
As shown in \Cref{fig:cloud-platform-schedprofile-results}(d), when using EEVDF with a 250\,Hz timer, the CPU time obtained often exceeds the configured quota, though it is slightly better than CFS with less overrun.
Raising the timer frequency to 1000\,Hz significantly mitigates the overrun issue.
However, even with higher timer frequencies, the fundamental overallocation problem still exists. Whenever required CPU time falls below the quota, overallocation cannot be avoided, regardless of scheduler or timer settings.

\textbf{\textit{Implications}}:
Overrun and overallocation are widespread on public serverless platforms.
However, providers can absorb this under-accounted resource usage through currently high invocation fees and coarse billing granularity (rounding up), as discussed in \S\ref{sec:high-invoke-fee-and-rounding}.
For example, a GCP function configured with 0.5\,vCPUs and 512\,MB memory can potentially consume 100\% CPU within 50\,ms, but GCP will round its billable wall-clock time up to 100\,ms plus a high invocation fee equivalent to 30.19\,ms.
Also, we tested a user-side exploit on AWS Lambda.
We implement an intermittent execution framework and decompose a long function (the video-processing application from SeBS~\cite{copik2021sebs}) into a sequence of short bursts, each falling within the quota.
We could reduce billable memory GB-seconds by $66.7\%$ on average (calculated over 100 data points).
However, because AWS charges a fixed invocation fee, our actual bill increased by $76.7\%$.
In other words, providers that plan to eliminate invocation fees and coarse billing granularity should account for these overallocation effects.

Besides billing, overallocation has clear performance impacts as shown in \Cref{fig:pyaes-overallocation}. 
Users can experience high jitters when vCPU allocations are near quantization boundaries.
Existing function-rightsizing tools~\cite{eismann2021sizeless,lin2022fine,moghimi2023parrotfish} are agnostic to the quantization effect we described. 
However, they should be able to capture this effect if equipped with fine-grained, data-driven search.
For the first time, we reveal the interplay between scheduling, performance, and billing that these frameworks implicitly use, potentially unlocking more optimal rightsizing strategies.

One potential way to address overrun and overallocation within the serverless computing context is to adopt an event-driven quota enforcement mechanism instead of periodic polling mechanisms based on periodic timers/ticks~\cite{tsafrir2007secretly}. 
For example, one-shot timers that expire upon a function process exhausting its bandwidth control quota may be set to trigger an immediate throttle and reschedule.
Also, per-task timers can be set to fire after a short, adaptive time (e.g., depending on the global bandwidth control period, overhead tolerance, accuracy requirements, and predicted task duration) to enforce more frequent and accurate CPU time accounting for short-lived tasks with fractional vCPU allocations.
In addition, BPF programs can be attached to the scheduler (e.g., through \texttt{sched\_ext}~\cite{web:schedext}) to selectively apply fine-grained quota enforcement to shorter functions that are more susceptible to overallocation.

\section{Discussions}
\label{sec:discussions}
\noindent\textbf{\textit{Relative contributions of each cost-related component}}: 
In this work, we chose not to quantify the relative contribution of each cost-related component since such numerical breakdowns are highly dependent on context-specific factors.
These factors include workload characteristics (e.g., traffic patterns, execution durations, and resource demands), user configurations (e.g., concurrency settings, provisioned resources~\cite{web:aws-reserved-concurrency}, and subscription plans~\cite{web:azure-consumption-billing}), and provider-specific policies (e.g., ARM CPU and committed use discounts and free tiers~\cite{web:aws-arm-discounts, web:lambda-pricing}), which vary across applications and providers.
Therefore, any numerical breakdown would not be broadly applicable.
Instead, our approach decomposes the inherent sources of cost inefficiencies and presents a systematic analysis framework across multiple abstraction layers from user-facing billing models to OS scheduling, which enables practitioners to measure and rank cost drivers within their own context.

\noindent\textbf{\textit{Actionables for serverless users}}: 
Our findings lead to several actionable recommendations for reducing serverless costs. First, users can conduct trace-based analysis to pick an appropriate platform whose cost drivers, such as billing practices (\S\ref{sec:billing-practices} and \Cref{tab:severless-platforms-billing}), concurrency modes (\S\ref{sec:serving-model}), serving architectures (\S\ref{sec:serving-arch-overhead}), keep-alive patterns (\S\ref{sec:ka-duration}), and scheduling granularity (\S\ref{sec:scheduling-profiling} and \Cref{tab:severless-platform-profiling}), best match their workload. 
Depending on the cost breakdown, users may consider merging similar functions to lower invocation fees, decomposing functions to better utilize resources, or configuring always-ready instances to avoid cold starts~\cite{baldini2017serverless, yu2020characterizing, mahgoub2022orion, web:aws-reserved-concurrency}.
Also, users should be wary of serverless concurrency models and tune control knobs for resources and scaling to avoid the dual penalty of slowdowns and higher bills (\S\ref{sec:serving-model}).
Furthermore, it is a good practice to tune workload resource demands and fractional vCPU allocations to avoid performance jitters due to coarse OS scheduling granularity (i.e., quantization jumps shown in \Cref{fig:pyaes-overallocation}).
Lastly, serverless users may also consider the possibility of running background tasks during keep-alive periods (\S\ref{sec:ka-duration} and \Cref{tab:ka-phase-behavior}).

\section{Related Work}
\label{sec:related_work}
\textbf{\textit{(1) External characterization of serverless systems}}: 
Numerous studies characterized serverless systems from the users' perspective.
Some investigated billing practices and cost efficiency of serverless platforms~\cite{adzic2017serverless, baldini2017serverless,eivy2017wary,lee2018evaluation, yu2020characterizing,liu2023demystifying,cvetkovic2023understanding}.
However, none offered a holistic top-down analysis like our study on how today’s serverless billing practices, request patterns, architectural overheads, and OS scheduling translate to inflated user costs.
Other works performed cross-platform characterizations of resource allocation patterns and performance variations of serverless offerings~\cite{wang2018peeking,kim2019functionbench, yu2020characterizing, copik2021sebs, wen2025scope, wen2025unveiling}.
They did not capture some critical factors that greatly impact serverless costs, such as concurrency models, details of resource allocation during keep-alive, and OS scheduling granularity. 
Moreover, some of their measurements, such as default keep-alive durations and serving architectures, are now outdated due to the rapid evolution of serverless platforms.

\noindent\textbf{\textit{(2) Characterization of production serverless workloads}}: Several provider-led studies characterized serverless workloads running within their systems~\cite{shahrad2020serverless, wang2021faasnet,joosen2023does, joosen2025serverless,zhang2021faster, mahgoub2022wisefuse}.
These provided valuable insights and enabled the trace-based analysis in parts of our study (e.g., quantifying inflated billable resources).
However, none of these studies delivers an in-depth analysis that correlates the internal platform characteristics (e.g., overheads of architectures and scheduler settings) with the high costs experienced by serverless users.
Our work fills this gap by holistically analyzing billing models and practices and measuring architectural overheads and OS scheduling effects, and connecting them to cost implications.

\noindent\textbf{\textit{(3) Open-source serverless solutions}}, such as Knative~\cite{web:knative}, AWS Runtime API~\cite{web:aws-golang-runtimeapi}, \texttt{workerd}~\cite{web:workerd}, and Azure Functions Host~\cite{web:azure-function-host}, enabled us to demystify overheads hidden in modern serverless serving infrastructures, which have not been reported before.

\noindent\textbf{\textit{(4) Serverless cost-efficiency optimization}}: There have been recent studies that leverage dynamic or cooperative scheduling~\cite{zhao2024serverless, cao2025making, pei2024litmus,fu2024alps, fu2022sfs}, adaptive overcommitment~\cite{tian2022owl,li2023golgi}, and new billing models~\cite{liu2023demystifying,pei2024litmus} to improve the resource utilization or cost-efficiency of serverless systems, effectively reducing costs.
These studies are orthogonal to our work, which, for the first time, comprehensively characterizes the interplay of various factors affecting performance and cost in a top-down manner from user-facing billing models to OS scheduling on major production serverless systems and holistically demystifies the high costs of serverless.

\section{Conclusion}
\label{sec:conclusion}
For the first time, we holistically demystify the high cost of serverless by conducting a comprehensive characterization of driving factors in a top-down manner from user-facing billing models, through architectural overheads, and finally to OS scheduling.
We provide novel insights into how current billing practices, request serving architectures, keep-alive resource allocation behaviors, and OS scheduling granularity contribute to the inflated costs of serverless.
These insights spark future directions for serverless practitioners on optimizing the cost-efficiency of serverless systems.

\begin{acks}
We thank the anonymous reviewers, and specially our shepherd, Andrea Arpaci-Dusseau, for helping us improve the paper.
We also thank Alain Tchana for his feedback.
This work was supported by the Natural Sciences and Engineering Research Council of Canada (NSERC), through research grants RGPIN-2021-03714 and DGECR-2021-00462, the Canada Graduate Scholarship (CGS D) program, and the Undergraduate Student Research Awards (USRA) program.
This work was also supported in part by the Institute for Computing, Information and Cognitive Systems (ICICS) at UBC.
Cloud resources from the Digital Research Alliance of Canada (RAS and RAC allocations) facilitated our research.
\end{acks}

\bibliographystyle{ACM-Reference-Format}
\bibliography{bibliography}

\end{document}